\documentclass[aps,%
twocolumn,%
floatfix,%
pra,%
amsfonts,
groupedaddress,
superscriptaddress]{revtex4-2}%
\usepackage[utf8]{inputenc}
\usepackage[T1]{fontenc}
\usepackage{mathrsfs}       
\usepackage{mathtools}
\usepackage{amsmath}
\usepackage{amsfonts}       
\usepackage[titletoc,title]{appendix}
\usepackage{algorithm,algorithmic}

\usepackage{bbm}            
\usepackage{bm}             
\usepackage{amsthm}         
\usepackage{makecell}
\usepackage[dvipsnames]{xcolor}
\definecolor{beamer@blendedblue}{rgb}{0.2,0.2,0.7}
\usepackage{times}
\usepackage{caption}
\usepackage{subfigure}
\usepackage{booktabs} 

\usepackage[colorlinks=true,%
linkcolor=beamer@blendedblue,%
bookmarks=true,%
breaklinks=true,%
filecolor=beamer@blendedblue,%
anchorcolor=yellow,%
citecolor=beamer@blendedblue,%
urlcolor=beamer@blendedblue,%
pdfauthor={Congcong Zheng, Xutao Yu, and Kun Wang},%
pdfsubject={Cross-Platform Comparison of Arbitrary Quantum Processes},%
CJKbookmarks=true]{hyperref}

\usepackage{float}    

\mathchardef\ordinarycolon\mathcode`\:
\mathcode`\:=\string"8000
\def\vcentcolon{\mathrel{\mathop\ordinarycolon}}
\begingroup \catcode`\:=\active
  \lowercase{\endgroup
  \let :\vcentcolon
  }

\DeclareFontFamily{U}{mathx}{\hyphenchar\font45}
\DeclareFontShape{U}{mathx}{m}{n}{<-> mathx10}{}
\DeclareSymbolFont{mathx}{U}{mathx}{m}{n}
\DeclareMathAccent{\widebar}{0}{mathx}{"73}

\newcommand{\ket}[1]{\left\vert{#1}\right\rangle}

\newcommand{\ketbra}[2]{\vert{#1}\rangle\!\langle{#2}\vert}

\DeclareMathOperator{\tr}{Tr}  
\newcommand{\1}{\mathbbm{1}}
\newcommand{\ox}{\otimes}

\makeatletter
\newsavebox{\@brx}
\newcommand{\llangle}[1][]{\savebox{\@brx}{\(\m@th{#1\langle}\)}%
  \mathopen{\copy\@brx\kern-0.5\wd\@brx\usebox{\@brx}}}
\newcommand{\rrangle}[1][]{\savebox{\@brx}{\(\m@th{#1\rangle}\)}%
  \mathclose{\copy\@brx\kern-0.5\wd\@brx\usebox{\@brx}}}
\makeatother

\newcommand*{\cD}{\mathcal{D}}
\newcommand*{\cE}{\mathcal{E}}

\newcommand*{\cH}{\mathcal{H}}
\newcommand*{\cI}{\mathcal{I}}

\newcommand*{\cS}{\mathcal{S}}

\newcommand*{\cU}{\mathcal{U}}

\newcommand*{\cX}{\mathcal{X}}

\begin{document}

\title{{Cross-Platform Comparison of Arbitrary Quantum Processes}}

\author{Congcong Zheng}
\affiliation{State Key Lab of Millimeter Waves, Southeast University, Nanjing 211189, China}
\affiliation{Frontiers Science Center for Mobile Information Communication and Security, Southeast University, Nanjing 210096, People's Republic of China}

\author{Xutao Yu}
\email{yuxutao@seu.edu.cn}
\affiliation{State Key Lab of Millimeter Waves, Southeast University, Nanjing 211189, China}
\affiliation{Frontiers Science Center for Mobile Information Communication and Security, Southeast University, Nanjing 210096, People's Republic of China}
\affiliation{Purple Mountain Laboratories, Nanjing 211111, People's Republic of China}

\author{Kun Wang}
\email{wangkun28@baidu.com}
\affiliation{Institute for Quantum Computing, Baidu Research, Beijing 100193, China}

\begin{abstract}

In this work, we present a protocol for comparing the performance of arbitrary quantum processes 
executed on spatially or temporally disparate quantum platforms using Local Operations and Classical Communication (LOCC). 
The protocol involves sampling local unitary operators, 
which are then communicated to each platform via classical communication
to construct quantum state preparation and measurement circuits.
Subsequently, the local unitary operators are implemented on each platform, 
resulting in the generation of probability distributions of measurement outcomes. 
The max process fidelity is estimated from the probability distributions,
which ultimately quantifies the relative performance of the quantum processes.
Furthermore, we demonstrate that this protocol can be adapted for quantum process tomography.
We apply the protocol to compare the performance of five quantum devices from IBM and the 
"Qianshi" quantum computer from Baidu via the cloud. 
Remarkably, the experimental results reveal that the protocol can accurately compare 
the performance of the quantum processes implemented on different quantum computers,
requiring significantly fewer measurements than those needed for full quantum process tomography.
We view our work as a catalyst for collaborative efforts in cross-platform comparison of quantum computers.

\end{abstract}

\date{\today}
\maketitle


\section{Introduction}

As the field of quantum computing and quantum information gains traction, 
an increasing number of manufacturers are entering the market, producing their own quantum computers. 
However, the current generation of noisy intermediate-scale quantum (NISQ) computers, 
despite their potential, are still hindered by quantum noise~\cite{preskill2018quantum}. 
A great challenge is how to compare the performance of the quantum computers fabricated by 
different manufacturers and located in different laboratories, termed as \textit{cross-platform comparison}.
This task is especially relevant when we move towards regimes where comparing to classical simulations becomes computationally challenging, 
and therefore a direct comparison of quantum computers is necessary. 

A standard method to achieve cross-platform comparison is the quantum tomography~\cite{o2016efficient}, 
in which we first reconstruct the full information of quantum computers under investigation,
and then estimate their relative fidelity from the obtained matrices. 
However, quantum tomography is known to be time consuming and computationally difficult; 
even learning a few-qubit quantum state is already experimentally challenging~\cite{haffner2005scalable,carolan2014experimental}.
A more efficient way is to estimate the fidelity of the quantum computer without resorting to the full information.
Indeed, a variety of estimation and verification tools~\cite{eisert2020quantum,kliesch2021theory}, 
such as fidelity estimation~\cite{hofmann2005complementary,bendersky2008selective,flammia2011direct,da2011practical,reich2013optimal,greenaway2021efficient} 
and quantum verification~\cite{pallister2018optimal,Zhu2019EfficientPRL,Wang2019Optimal,yu2022statistical}, have been developed along this way.
However, these methods assume that one can access a known and theoretical target, usually simulated by classical computers.
They quickly become inaccessible for quantum computers containing several hundreds or even thousands of highly entangled qubits, 
due to the intrinsic time complexity of classical simulation.

Recently, Elben \textit{et al.}~\cite{elben2020cross} proposed the first cross-platform protocol 
for estimating the fidelity of quantum states, which are possibly generated by
spatially and temporally separated quantum computers.
This protocol requires only local measurements in randomized product bases and classical communication between quantum computers.
Numerical simulation shows that it consumes significantly fewer measurements than full quantum state tomography.
It is expected be applicable in state-of-the-art quantum computers consisting of a few tens of qubits.
Later on, Kn\"{o}rzer \textit{et al.}~\cite{knorzer2022cross} extended Elben's protocol to cross-platform
comparison of quantum networks, assuming the existence of quantum links that can teleport quantum states.
Nevertheless, a quantum link transferring quantum states of many qubits with high accuracy 
between two distant quantum computers is far from reach in the near future.

In this work, by elaborating the core idea of~\cite{elben2020cross}, 
we present a novel protocol for cross-platform comparing spatially and temporally separated quantum processes. The protocol uses only single-qubit unitary gates and classical communication between quantum computers, without requiring quantum links or ancilla qubits. This approach allows for accurate estimation of the performance of quantum devices manufactured in separate laboratories and companies using different technologies. Furthermore, the protocol can be used to monitor the stable function of target quantum computers over time. We apply the protocol to compare the performance of five quantum devices from IBM and the "Qianshi" quantum computer from Baidu via the cloud. Our experimental results reveal that our protocol can accurately compare the performance of arbitrary quantum processes. Although the sample complexity of our protocol still scales exponentially with the number of qubits, it has a significantly smaller exponent factor compared with that of quantum process tomography. Overall, our protocol serves as a novel application of the powerful randomized measurement toolbox~\cite{Elben2022}.

The rest of the paper is organized as follows. 
Section~\ref{sec:Preliminaries} reviews the cross-platform quantum state comparison protocol in~\cite{elben2020cross}
and introduces the quantum process performance metric.
Section~\ref{sec:theory} elaborates the main result, a new protocol for cross-platform comparing arbitrary quantum processes.
Particularly, we summarize the similarities and differences between our protocol and that of~\cite{knorzer2022cross}. 
Section~\ref{sec:experimental-results} reports a thorough experimental cross-platform comparison
on spatially and temporally separated quantum computers, along with a comprehensive investigation of the experimental data.
The Appendices summarize technical details of the main text. 


\section{Preliminaries}\label{sec:Preliminaries}


\subsection{Cross-platform comparison of quantum states}\label{sec: state fidelity}

In quantum information, fidelity is an important metric that is widely used to 
characterize the closeness between quantum states. 
There are many different proposals for the definition of state fidelity~\cite{liang2019quantum}.
In this work, we will concentrate on the \emph{max fidelity}, formally defined as \cite{liang2019quantum,elben2020cross}
\begin{align}
  F_{\max}(\rho_1, \rho_2) := \frac{\tr[\rho_1\rho_2]}{\max\{\tr[\rho_1^2], \tr[\rho_2^2]\}}, \label{eq: definition of max state fidelity}
\end{align}
where $\rho_i$ is an $n$-qubit quantum state produced by the quantum computer, $i=1,2$. 

Elben \textit{et al.}~\cite{elben2020cross} proposed a randomized measurement protocol to estimate $F_{\max}$,
which functions as follows.
First, we construct an $n$-qubit unitary $U = \bigotimes_{k=1}^n U_k$, 
where each $U_k$ is identically and independently sampled from a single-qubit set $\cX_2$ 
satisfying unitary $2$-design~\cite{dankert2009exact, gross2007evenly}.
This information will be classically communicated to the quantum computers, possibly spatially or temporally separated,
that produce the quantum states $\rho_1$ and $\rho_2$, respectively.
Then, each quantum computer executes the unitary $U$, performs a computational basis measurement, and records the measurement outcome $\bm{s}$.
Repeating the above procedure for fixed $U$ a number of times, 
we are able to obtain two probability distributions over the outcomes of the form $\Pr_U^{(1)}, \Pr_U^{(2)}$,
where the superscript $i$ represents that the distribution is obtained from quantum state $\rho_i$.
Next, we repeat the whole procedure for many different random unitaries $U$,
yielding a set of probability distributions $\{\Pr_U^{(1)}, \Pr_U^{(2)}\}_U$.
From the experimental data, we estimate the overlap between $\rho_i$ and $\rho_j$ as~\cite{elben2020cross}
\begin{align}
  \tr[\rho_i\rho_j] = 2^n \sum_{\bm{s}, \bm{s}'\in\{0,1\}^{n}} 
  (-2)^{-\cD[\bm{s}, \bm{s}']}\overline{{\rm Pr}_U^{(i)}[\bm{s}]{\rm Pr}_U^{(j)}[\bm{s}']}, \label{eq: cpe for state}
\end{align}
where $\overline{\cdots}$ denotes the ensemble average over the sampled unitaries $U$ 
and $\cD[\bm{s}, \bm{s}']$ denotes the hamming distance between two bitstrings $\bm{s}$ and $\textbf{s}'$. 
Specially, $\tr[\rho_1\rho_2]$ can be estimated from~\eqref{eq: cpe for state} by setting $i=1$ and $j=2$, 
whereas the purities $\tr[\rho_1^2]$ and $\tr[\rho_2^2]$ can be obtained by setting $i=j=1$ and $i=j=2$, respectively.
Using the above estimated quantities, 
we successfully compute the max fidelity $F_{\max}(\rho_1, \rho_2)$.

Using experimental data from~\cite{brydges2019probing},
Elben \textit{et al.} showcased the experiment-theory fidelities
and experiment-experiment fidelities of highly entangled quantum states prepared 
via quench dynamics in a trapped ion quantum simulator as a proof of principle~\cite{elben2020cross}.
Recently, Zhu \textit{et al.} reported thorough cross-platform comparison of 
quantum states in four ion-trap and five superconducting quantum platforms, with detailed analysis of the results and 
an intriguing machine learning approach to explore the data~\cite{zhu2022cross}.


\subsection{Quantum process performance metric}

A quantum process, also known as a quantum operation or a quantum channel, 
is a mathematical description of the evolution of a quantum system. 
It is mathematically formulated as a completely positive and trace-preserving (CPTP) linear 
map on the quantum states~\cite{watrous2018theory}. 
The Choi-Jamio\l{}kowski isomorphism provides a unique way to represent quantum processes as quantum states in a larger Hilbert space.
Formally, the Choi state of an $n$-qubit quantum process $\cE$ is defined as~\cite{jamiolkowski1972linear}
\begin{align}
  \eta_{\cE} := (\cI\ox\cE )\ketbra{\psi_+}{\psi_+}, \label{eq: definition of choi state}
\end{align}
where $\cI$ is the identity channel and $\ket{\psi_+} := 1/\sqrt{2^n}\sum \ket{ii}$ is 
a maximally entangled state of a bipartite quantum system composed of two $n$-qubit subsystems.

One lesson we can learn from the cross-platform state comparison protocol is that, 
we must choose a process metric before comparing two quantum processes.
Gilchrist \textit{et al.}~\cite{gilchrist2005distance} have introduced a systematic way to 
generalize a metric originally defined on quantum states to a corresponding metric on quantum processes,
utilizing the Choi-Jamio\l{}kowski isomorphism.
Specifically, the \emph{max fidelity} between two $n$-qubit quantum 
processes $\cE_1$ and $\cE_2$, implemented on different quantum platforms, is defined as
\begin{align}
      F_{\max}(\cE_1, \cE_2) := F_{\max}(\eta_{\cE_1}, \eta_{\cE_2}), \label{eq:cp process fidelity}
\end{align}
where $\eta_\cE$ is the Choi state of quantum process $\cE$.
This metric fulfills the axioms for quantum process fidelities following the argument in~\cite{gilchrist2005distance}.
It is reasonable to believe that, at least to some extent, this metric reveals that
the quantum processes have implemented the same quantum evolution.

In the following, we propose an experimentally efficient protocol to estimate this metric.
This protocol makes use of only single-qubit unitaries and classical communication,
thus can be executed in spatially and temporally separated quantum devices.
This enables cross-platform comparison of arbitrary quantum processes.


\section{Cross-Platform Comparison}\label{sec:theory}

In this section, we first provide a simple example to illustrate the necessity of cross-platform comparison.
Then, we introduce a protocol for estimating the max process fidelity that is conceptually straightforward yet experimentally challenging.
Next, we propose a modification to the protocol that employs randomized input states and provide a detailed explanation of the approach. 
Furthermore, we demonstrate that our protocol can be extended to accomplish full process tomography.
Our protocol is motivated by the observation that 
even identical quantum computers cannot produce identical outcomes on each run due to the intrinsic randomness of quantum mechanics, 
but they do generate identical probability distributions from a statistical perspective.

Cross-platform comparison of quantum computers is essential for at least two reasons.
Firstly, comparing the actual implementation with an idealized theoretical simulation can be challenging, 
as classical simulations become computationally demanding with an increasing number of qubits. 
Secondly, due to the presence of varying forms of quantum noise across different quantum platforms, 
the actual implementation of quantum processes can vary significantly, 
even if they maintain the same process fidelity with respect to the ideal target.
To illustrate this point, consider the following example.
Suppose Alice has a superconducting quantum computer and Bob has a trapped-ion quantum computer.
They implement the single-qubit Hadamard gate $\cH(\rho)=H\rho H^\dagger$ on their respective quantum computers.
However, Alice's implementation $\cE_1$ suffers from the depolarizing noise, 
yielding $\cE_1(\rho)=(1-p_1) H\rho H^\dagger + p_1\1/2$, where $p_1=7/30$.
On the other hand, Bob's implementation $\cE_2$ suffers from the dephasing noise,
such that $\cE_2(\rho)=(1-p_2)H\rho H^\dagger + p_2{\rm\Delta(\rho)}$, where $p_2=1/5$ and $\Delta(\cdot)$ is the dephasing operation.
After simple calculations, we obtain $F_{\max}(\cE_1,\cH)=F_{\max}(\cE_2,\cH)\approx 0.808$ and $F_{\max}(\cE_1,\cE_2)\approx 0.978$.
Despite $\cE_1$ and $\cE_2$ having the same fidelity level when compared to the ideal target $\cH$, 
discernible difference exists between them. 
Therefore, solely comparing the fidelity of a quantum process to an ideal reference is insufficient, 
and a direct comparison between quantum processes is warranted.

\subsection{Ancilla-assisted cross-platform comparison}\label{sec:ancilla-free}

\begin{figure}[!htbp]
	\centering
	\includegraphics[width=0.46\textwidth]{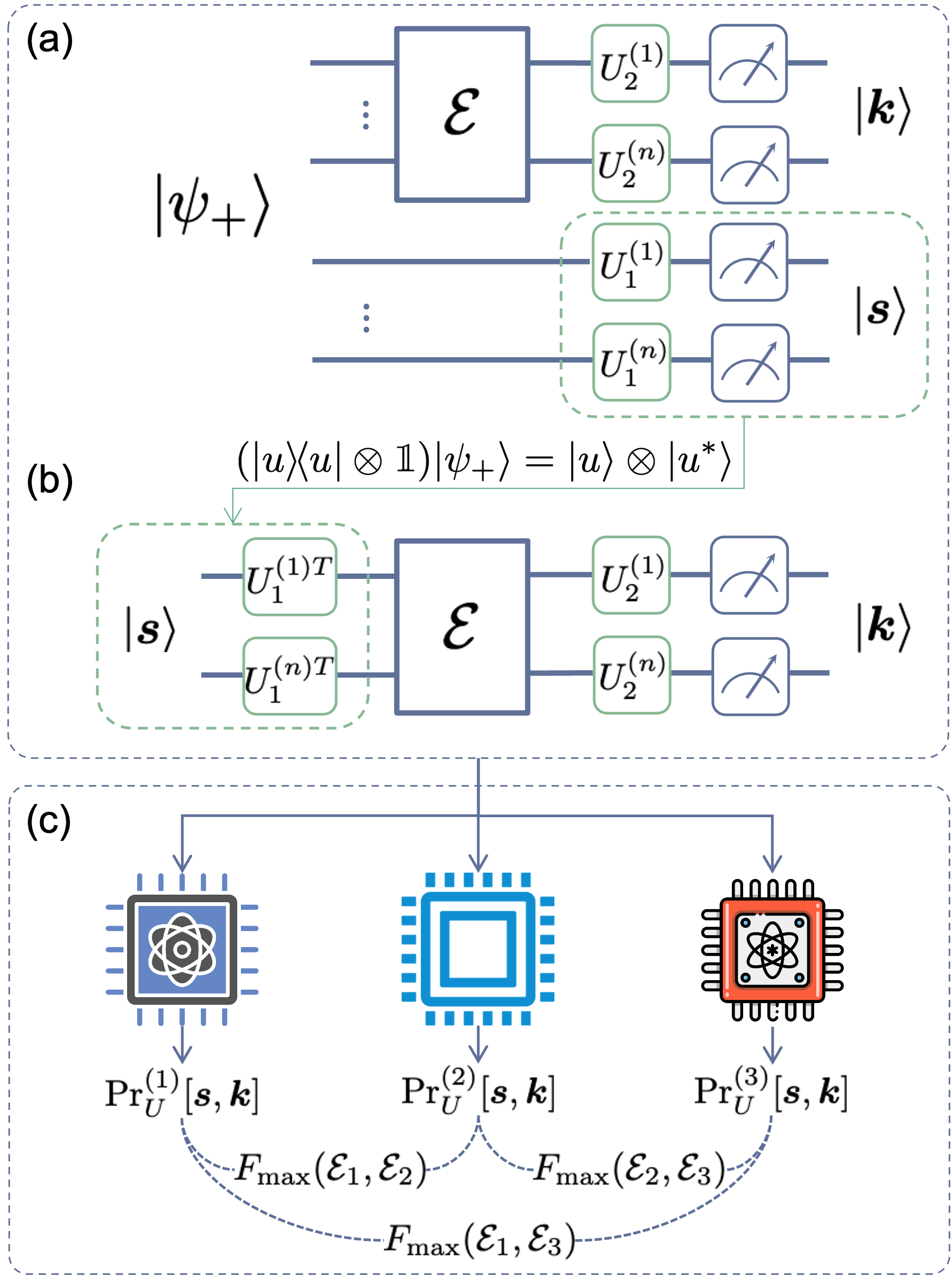}
	\caption{\raggedright 
            Two protocols to estimate the max process fidelity $F_{\max}$ between 
            quantum processes implemented on different quantum platforms. 
            (a) \emph{Ancilla-assisted protocol}: 
            Prepare the maximally entangled state, 
            execute the target quantum process, and
            perform the randomized measurements given by $\otimes_{k=1}^n U_1^{(k)}\otimes\otimes_{k=1}^n U_2^{(k)}$. 
            (b) \emph{Ancilla-free protocol}: 
            Randomly sample a computational basis $\ket{\bm{s}}$,
            execute the unitaries $\otimes_{k=1}^n U_1^{(k)T}$, 
            execute the target quantum process, and
            perform the randomized measurements given by $\otimes_{k=1}^n U_2^{(k)}$.
            (c) Run the quantum circuits constructed in (a) or (b) on platform $\cS_i$
                to obtain the probability distribution $\Pr_U^{(i)}[\bm{s}, \bm{k}]$.
                The max process fidelity $F_{\max}(\cE_i, \cE_j)$ is inferred from the probability distributions (see text).}
	\label{fig: framework}
\end{figure}

In this section, we recover a conceptually simple approach
for estimating the max process fidelity defined in Eq.~\eqref{eq:cp process fidelity},
which was recently proposed in~\cite{knorzer2022cross}.
The key observation is that this fidelity can be seen as the max state fidelity 
between the Choi states of the corresponding quantum processes. 
To construct the Choi state of the $n$-qubit quantum process $\cE$, 
we need to introduce an additional $n$-qubit clean auxiliary system.
Using the auxiliary system, we prepare a $2n$-qubit maximally entangled state $\vert \psi_+\rangle$ 
and apply $\cE$ to half of the whole system, which successfully prepares the Choi state of $\cE$. 
We can then estimate the max state fidelity using the procedure introduced in Section~\ref{sec: state fidelity}. 
The complete protocol is illustrated in Figure~\ref{fig: framework}(a)-(c).

We refer to this protocol as the \emph{ancilla-assisted cross-platform comparison} 
because it requires additional clean ancilla qubits to prepare the Choi state of the quantum process. 
To perform this protocol, a maximally entangled state is required as input, 
resulting in a two-fold overhead when comparing $2n$-qubit states instead of $n$-qubit states. 
Consequently, this protocol may not be practical in scenarios with limited quantum computing resources. 
Furthermore, preparing high-fidelity maximally entangled states can be experimentally challenging, 
which may negatively impact the accuracy of the protocol.

\subsection{Ancilla-free cross-platform comparison}\label{sec: ancilla free cross-platform comparison}

To overcome the limitations of the ancilla-assisted protocol, 
we propose an efficient and ancilla-free approach for estimating the max process fidelity.
Our protocol does not require any additional qubits or the preparation of maximally entangled states.
The key observation is that the auxiliary system in the ancilla-assisted protocol only needs to perform randomized measurements.
After the measurement, the auxiliary system collapses to one eigenstate of the sampled measurement operator. 
Based on the identity $(\ketbra{u}{u}\ox\1)\ket{\psi_+} = \ket{u}\ox\ket{u^*}$, where $\1$ is the identity matrix,
and the deferred measurement principle~\cite{nielsen_chuang_2010}, we can eliminate the auxiliary system 
by preparing computational states and applying the transposed unitary operator on the main system.
Please refer to Appendix~\ref{appx: proof of cpe for process} for a detailed analysis.

We refer to the new protocol as the \emph{ancilla-free cross-platform comparison} and it works as follows.
We consider two $n$-qubit quantum processes $\cE_1$ and $\cE_2$ realized on different quantum platforms $\cS_1$ and $\cS_2$,
whose Choi states are $\eta_1$ and $\eta_2$, respectively. 
The protocol, illustrated in Figure~\ref{fig: framework}(b)-(c), 
consists of three main steps: sampling unitaries, running circuits, and post-processing.

\vspace{0.05in}
\textbf{Step 1. Sampling unitaries:} 
Construct two $n$-qubit unitaries $U_i = \bigotimes_{k=1}^n U_i^{(k)}$, $i=1,2$,
where each $U_i^{(k)}$ is identically and independently sampled from a single-qubit set $\cX_2$ satisfying unitary $2$-design.
The information of $U_i$ is then communicated to both platforms via classical communication. 

\vspace{0.05in}
\textbf{Step 2. Running circuits:} After receiving the information of the sampled unitaries, 
Each platform $\cS_i$ ($i=1,2$) initializes its quantum system to the computational states $\ket{\bm{s}}$ 
and applies the first unitary $U_1$ to $\ket{\bm{s}}$. 
Subsequently, $\cS_i$ implements the quantum process $\cE_i$ and applies the second unitary $U_2$. 
Finally, $\cS_i$ performs the projective measurement in the computational basis and obtains an outcome $\bm{k}$.
Repeating the above procedure many times, 
we obtain two probability distributions $\Pr_{K\vert\bm{s}, U_1,U_2}^{(1)}$ and $\Pr_{K\vert\bm{s}, U_1,U_2}^{(2)}$ 
over the measurement outcomes $\bm{k}$ for the fixed computational state $\ket{\bm{s}}$ and unitaries $U_1$ and $U_2$. 
By exhausting the computational states and repeatedly sampling the unitaries, 
we obtain two probability distributions $\Pr_{K,S\vert U_1,U_2}^{(i)}$ 
with respect to the sampled unitaries and computational state inputs.
For simplicity, we abbreviate $\Pr_{K,S|U_1,U_2}^{(i)}$ to $\Pr^{(i)}_{U}$. 

\vspace{0.05in}
\textbf{Step 3. Post-processing:} From the experimental data, 
we estimate the overlap between the Choi states $\eta_i$ and $\eta_j$ for $i,j=1,2$ as
\begin{align}
  \begin{split}
    \tr[ \eta_i\eta_j] = 
    4^n\sum_{\bm{s}, \bm{s}', \bm{k}, \bm{k}'\in\{0,1\}^n}
    & (-2)^{-\cD[\bm{s}, \bm{s}'] -\cD[\bm{k}, \bm{k}']} \\ 
    &\times\overline{{\rm Pr}^{(i)}_{U}[\bm{s},\bm{k}]{\rm Pr}^{(j)}_{U}[\bm{s}',\bm{k}']}.
  \end{split} \label{eq: analyzing result}
\end{align}
where $\overline{\cdots}$ denotes the ensemble average over the sampled unitaries $U_1$ and $U_2$.
This is proven in Appendix~\ref{appx: proof of cpe for process}.
By setting $i=1$ and $j=2$, we can estimate the overlap $\tr[\eta_1\eta_2]$ from the above equation, 
which is the second-order cross-correlation of the probabilities $\Pr^{(1)}_{U}$ and $\Pr^{(2)}_{U}$. 
We can obtain the purities $\tr[\eta_1^2]$ and $\tr[\eta_2^2]$ by setting $i=j=1$ and $i=j=2$, respectively. 
These are the second-order autocorrelations of the probabilities.
Using the estimated quantities, we compute the max process fidelity $F_{\max}(\cE_1, \cE_2)$ in Eq.~\eqref{eq:cp process fidelity}. 

There are several important points to note about our protocol. 
First, when classical simulation is available, the protocol can be used to compare the experimentally implemented process to the theoretical simulation, 
providing a useful tool for experiment-theory comparison. 
Second, our protocol can also estimate the process purity $\tr[\eta_{\cE}^2]$ of a quantum process $\cE$, 
which measures the extent to which $\cE$ preserves the purity of the quantum state. 
This is an important measure for characterizing quantum processes, and our protocol provides an efficient way to estimate it. 
Finally, it is worth noting that the definition of max process fidelity is not unique, and different approaches exist~\cite{gilchrist2005distance,korzekwa2018coherifying}. 
Our protocol, based on statistical correlations of randomized inputs and measurements, 
can be readily extended to any metric that depends solely on the process overlap $\tr[\eta_{\cE_1}\eta_{\cE_2}]$ 
and the process purities $\tr[\eta_{\cE_1}^2]$ and $\tr[\eta_{\cE_2}^2]$. 
This makes our protocol highly versatile and applicable to a wide range of quantum computing scenarios.

\subsection{Randomized quantum process tomography}

Here we argue that our protocol is applicable for full quantum process tomography. 
It is worth noting that in Ref.~\cite{elben2019statistical}, 
a method was proposed for performing full quantum state tomography using randomized measurements. 
For an $n$-qubit quantum process $\cE$, we can first construct the Choi state of $\cE$ and 
then use the proposed protocol to obtain the full information of the Choi state $\eta_\cE$. 
However, as previously mentioned, this method is not efficient and is impractical due to the imperfect preparation 
of maximally entangled states and the requirement for an additional $n$-qubit auxiliary system.

Likewise, we may use the randomized input states trick introduced in Section~\ref{sec: ancilla free cross-platform comparison}
to overcome the above issues. Specifically, based on the experimental data $\Pr_{U}$
collected in Section~\ref{sec: ancilla free cross-platform comparison},
the full information of an unknown $n$-qubit quantum process $\cE$ can be obtained via 
\begin{align}
  \begin{split}
    \eta_\cE = 
    4^n\sum_{\bm{s}, \bm{s}', \bm{k}, \bm{k}'\in\{0,1\}^n}
    & (-2)^{-\cD[\bm{s}, \bm{s}'] -\cD[\bm{k}, \bm{k}']} \\ 
    &\times\overline{{\rm Pr}_{U}[\bm{s},\bm{k}] U^\dagger\ketbra{\bm{s}'\bm{k}'}{\bm{s}'\bm{k}'}U}, 
  \end{split}
\end{align}
where $U = U_1 \ox U_2$ and $\overline{\cdots}$ denotes the ensemble average over the sampled unitaries $U_1$ and $U_2$ as before.
The is proven in Appendix~\ref{appx: proof of randomized qpt}. 


\begin{figure*}[!htbp]
  \centering
  \includegraphics[width=0.8\textwidth]{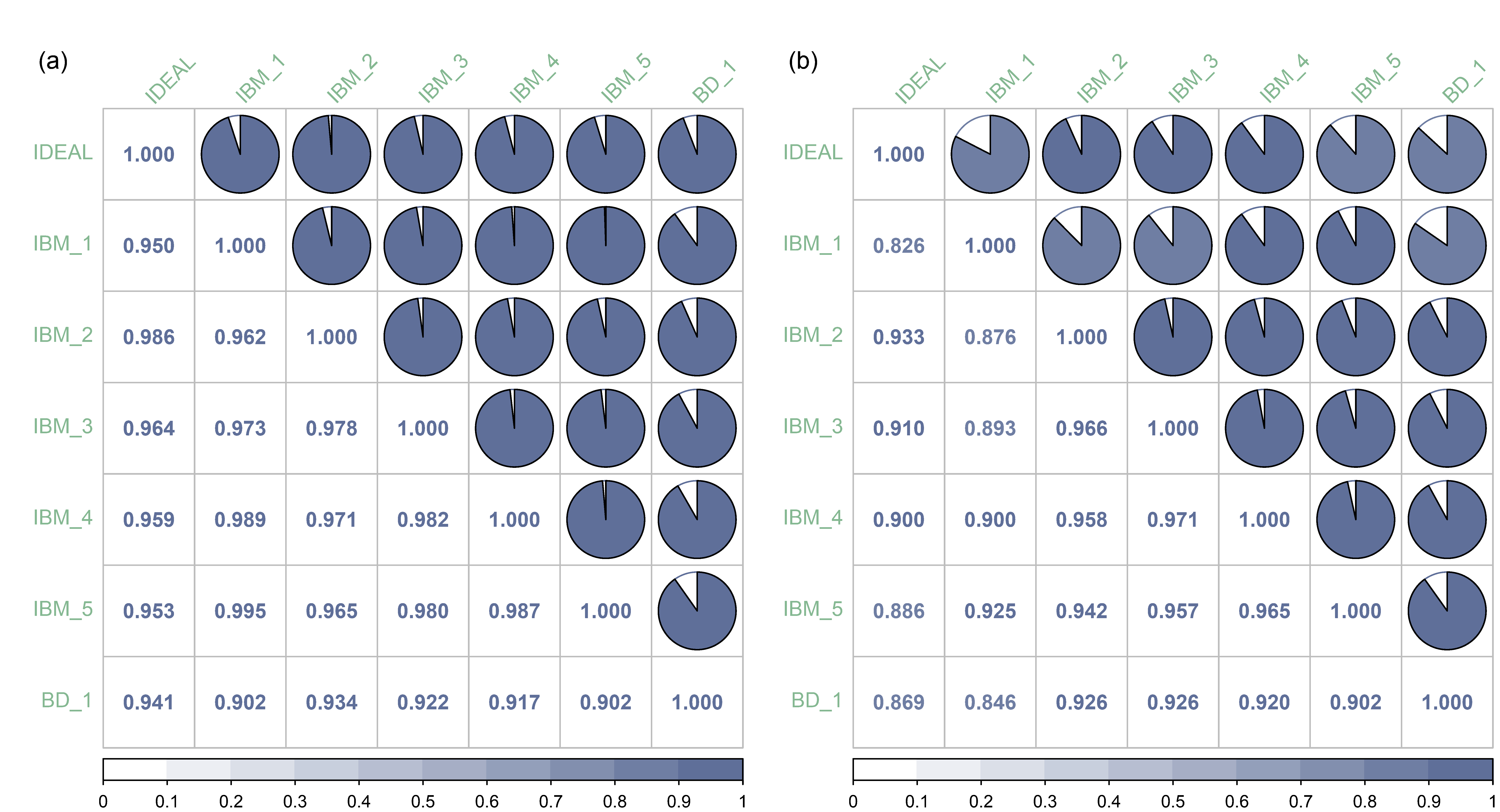}
  \caption{\raggedright
          The performance matrices for the single-qubit \texttt{H} and two-qubit \texttt{CNOT} gates
          generated from seven different quantum platforms.
          The entry in the $i$-th row and $j$-th column of the matrix represents the max process fidelity between platform-$i$ and platform-$j$. 
          The entries in the upper right corner are visualized in pie chart format.
          (a) The performance matrix of the \texttt{H} gate. 
          Each entry is inferred from $2^1 \cdot N_U = 20$ random circuits and each circuit is repeated $M_{\rm shots}=500$ times. 
          (b) The performance matrix of the \texttt{CNOT} gate. 
          Each entry is inferred from $2^2 \cdot N_U = 20$ random circuits and each circuit is repeated $M_{\rm shots}=500$ times.} 
  \label{fig: experiment-1}
\end{figure*}

\subsection{Comparison with previous works}

Kn\"{o}rzer \textit{et al.}~\cite{knorzer2022cross} have recently introduced a new set of protocols that 
enable pair-wise comparisons between distant nodes in a quantum network. The authors propose four cross-platform state 
comparison schemes as alternatives to Elben's protocol~\cite{elben2020cross}, 
each of which relies on the presence of quantum links. 
In addition to this, they present three protocols, referred to as M1, M2, and M3, which facilitate 
cross-platform comparisons of quantum processes, assuming that a cross-platform state comparison protocol is available.

We will now explain how our protocol differs from M1, M2, and M3.
While M1 involves an ancilla-assisted comparison protocol that
we have rephrased in Section~\ref{sec:ancilla-free}, our protocol does not rely on ancilla qubits. 
Similarly, M3 involves a series of entanglement tests that are fundamentally different from our protocol. 
Although our protocol and M2 share some similarities, such as the absence of ancilla qubits and 
the need to sample random unitaries and computational basis states, there are notable differences.
Specifically, our protocol only needs to sample from a \emph{single-qubit} unitary $2$-design, 
can accurately estimate the max fidelity, and can compare the performance of arbitrary quantum processes.
On the other hand, M2(i) estimates the average gate fidelity and requires sampling from a \emph{multi-qubit} unitary $2$-design,
which can be resource-intensive as the number of qubits increases.
M2(iii) is conceptually straightforward but can only estimate the ability of quantum processes to preserve quantum information in the computational basis. 
Additionally, M2(i) and M2(iii) are limited to comparing the performance of unitary quantum processes.


\begin{figure*}[!htbp]
  \centering
  \includegraphics[width=0.8\textwidth]{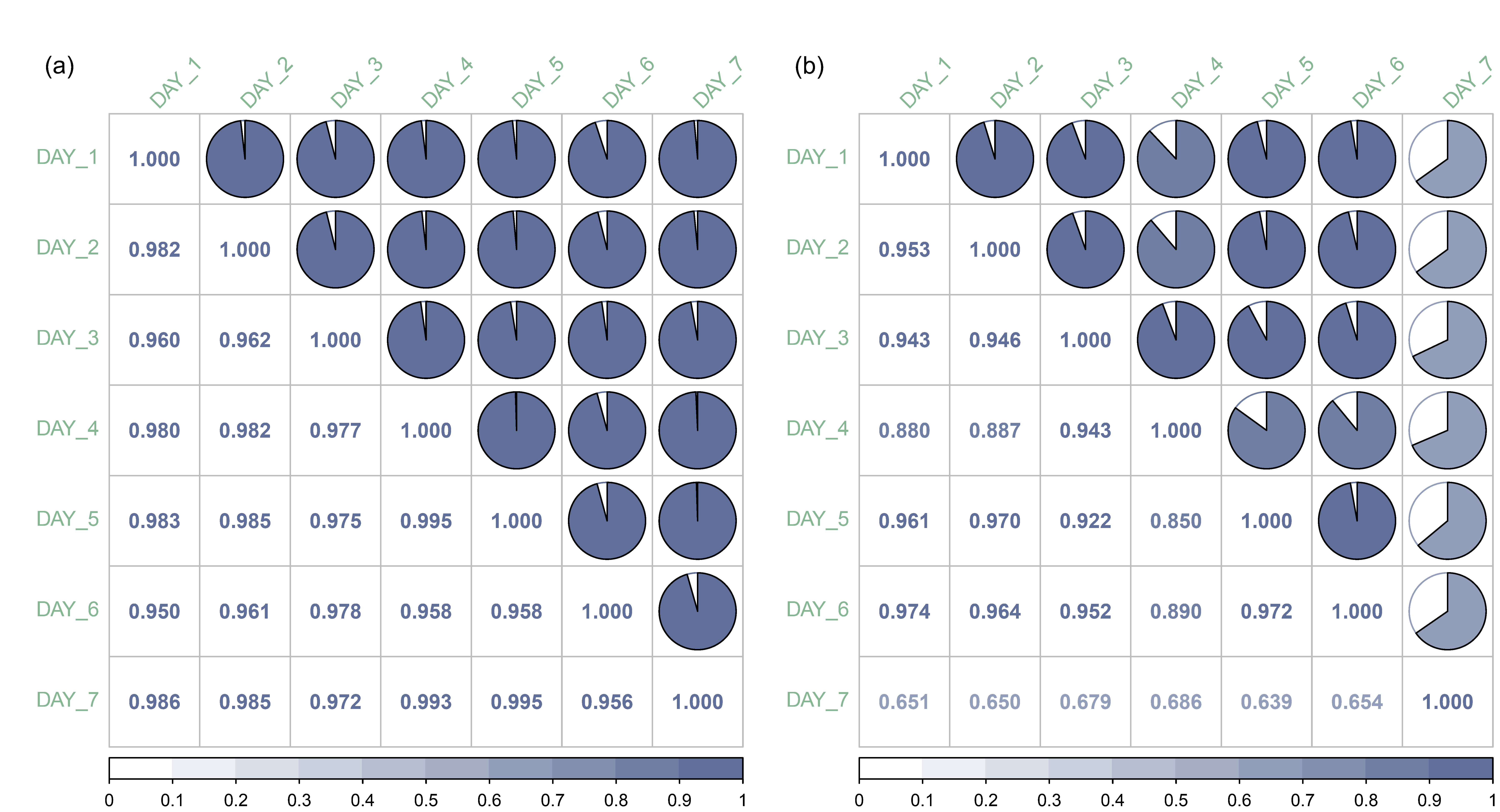}
  \caption{\raggedright
          The performance matrices of the single-qubit \texttt{H} and two-qubit \texttt{CNOT} gates
          generated from the daily data of Baidu's "Qianshi" quantum computer for one week.
          The entry in the $i$-th row and $j$-th column of the matrix represents the max process fidelity between platform-$i$ and platform-$j$. 
          The entries in the upper right corner are visualized in pie chart format.
          (a) The performance matrix of the \texttt{H} gate. 
          Each entry is inferred from $2^1 \cdot N_U = 20$ random circuits and each circuit is repeated $M_{\rm shots}=500$ times. 
          (b) The performance matrix of the \texttt{CNOT} gate. 
          Each entry is inferred from $2^2 \cdot N_U = 400$ random circuits and each circuit is repeated $M_{\rm shots}=500$ times.} 
  \label{fig: experiment-1-times}
\end{figure*}

\section{Experiments}\label{sec:experimental-results}

In this section, we report experimental results on cross-platform comparison of quantum processes
across various spatially and temporally separated quantum devices.
First, we demonstrate the efficacy of our protocol in comparing the \texttt{H} and \texttt{CNOT} gates
implemented on different platforms with their ideal counterparts obtained from classical simulation.
Next, we monitor the stability of the "Qianshi" quantum computer from Baidu over a week with our protocol.
Finally, we conduct an extensive numerical analysis to determine the expected number of experimental runs required to obtain reliable results.
All the experiments are conducted using the Quantum Error Processing toolkit developed on the Baidu Quantum Platform~\cite{QEP2023}.

\subsection{Comparing spatially separated quantum processes}\label{sec:spatially separated}

We utilize our ancilla-free cross-platform comparison protocol to assess the performance
of \texttt{H} and \texttt{CNOT} gates implemented on seven distinct platforms that are freely accessible to the public over the internet.
These platforms include six superconducting quantum computers, namely
\textit{ibmq\_quito} (IBM\_1), 
\textit{ibmq\_oslo} (IBM\_2), 
\textit{ibmq\_lima} (IBM\_3), 
\textit{ibm\_nairobi} (IBM\_4), 
\textit{ibmq\_manila} (IBM\_5), and 
\textit{baidu\_qianshi} (BD\_1), 
as well as the \textit{baidu ideal simulator} (IDEAL), which is intended for experiment-theory comparisons.

First of all, it is noteworthy that the random Pauli basis measurements $\{X,Y,Z\}$ are equivalent to randomized measurements 
with a single qubit Clifford group~\cite{zhu2022cross,huang2020predicting}.
The Clifford group is a unitary $2$-design group, and it can be employed to conduct complete process tomography of $n$-qubit quantum processes. 
This equivalence enables us to sample directly from the $3^n$ Pauli preparation and $3^n$ Pauli measurement unitaries in our experiments.

To begin with, we utilize our protocol to compare the performance of the single-qubit \texttt{H} gate across seven quantum platforms. 
To achieve this, we create $2^1\times N_U = 20$ random circuits and execute $M_{\rm shots} = 500$ projective measurements for each circuit on each platform.
Furthermore, we employ the same protocol to compare the performance of the \texttt{CNOT} gate implemented on these platforms. 
To accomplish this, we generate $2^2\times N_U = 400$ random circuits and performe $M_{\rm shots} = 500$ repetitions for each quantum circuit on each platform.
The performance matrices for the \texttt{H} and \texttt{CNOT} gates are presented in Figure~\ref{fig: experiment-1}. 

The experimental results make it clear that, while some quantum devices may achieve fidelities that are comparable to those of the ideal simulator, 
there remains a significant discrepancy between them. 
This emphasizes the importance of directly comparing the performance of quantum devices with each other, 
rather than relying solely on comparisons to an ideal simulator, as such comparisons may not be adequate.

\subsection{Comparing temporally separated quantum processes}

Our protocol is also useful for monitoring the stable performance of quantum devices. 
To this end, we employe the ancilla-free cross-platform comparison protocol to assess the stability of 
\texttt{H} and \texttt{CNOT} gates implemented on Baidu's "Qianshi" quantum computer (BD\_1) over the course of one week.
The experimental settings for the \texttt{H} and \texttt{CNOT} gates are identical to those used in the previous section. 
Specifically, for the single-qubit \texttt{H} gate, 
we create $2^1\times N_U = 20$ random circuits daily and execute $M_{\rm shots} = 500$ projective measurements for each circuit on "Qianshi".
For the two-qubit \texttt{CNOT} gate,
we create $2^2\times N_U = 400$ random circuits daily and execute $M_{\rm shots} = 500$ projective measurements for each circuit on "Qianshi".
The performance matrices of the single-qubit \texttt{H} and two-qubit \texttt{CNOT} gates 
generated from the daily data of "Qianshi" are shown in Figure~\ref{fig: experiment-1-times}. 

After analyzing the cross-platform fidelities presented in Figure~\ref{fig: experiment-1-times}, we discover several noteworthy features. 
First, we observe that the stability of the \texttt{H} gate is considerably higher than that of the \texttt{CNOT} gate on "Qianshi," 
which aligns with the expectation that two-qubit gates are harder to implement and maintain in a superconducting quantum computer than single-qubit gates.
Additionally, on the last day of the week (DAY\_7), there is a significant drop in the performance of the \texttt{CNOT} gate.
After consulting with researchers from Baidu's Quantum Computing Hardware Laboratory, 
it is determined that the instability is caused by the sudden halt of the dilution cooling system. 
After the system is restarted, all native quantum gates have to be re-calibrated to achieve optimal performance.
Furthermore, it is observed that the temperature variation had a negligible impact on the \texttt{H} gate.
This observation might be helpful for the experimenters to identify potential hardware issues.

\subsection{Scaling of the required number of experimental runs}

\begin{figure}[!htbp]
  \centering
  \includegraphics[width=0.4\textwidth]{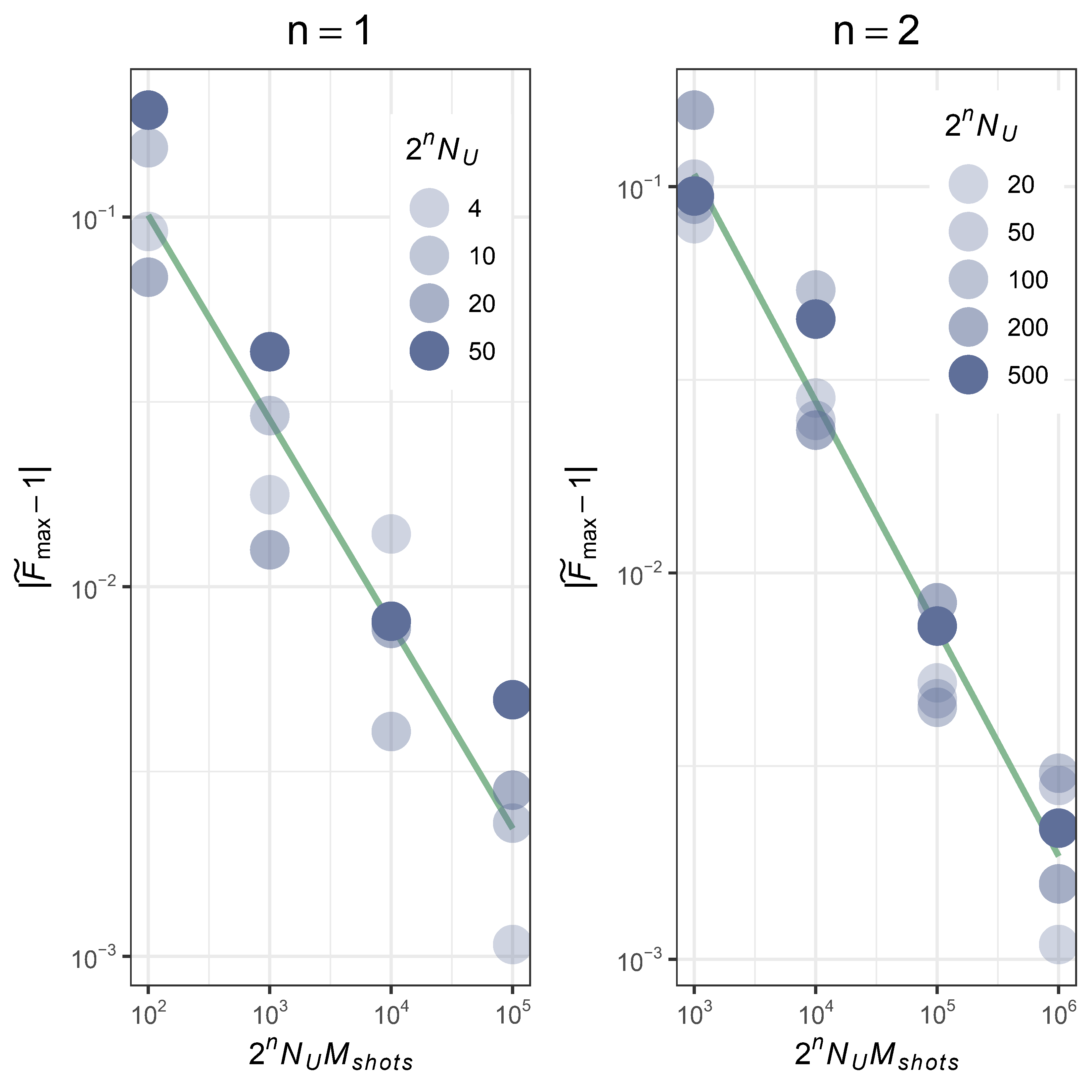}
  \caption{\raggedright
          Average statistical error $\vert\widetilde{F}_{\max} - 1\vert$
          as a function of the total number of experimental runs $2^n N_U M_{\rm shots}$.
          The target quantum process is taken to be the \texttt{H} gate for $n=1$ and the \texttt{CNOT} gate for $n=2$.
          The green lines obey $\sim 1/(2^n N_U M_{\rm shots})$ and are guides for the eye.
          The data is obtained via numerical simulation.}
  \label{fig: experiment-2}
\end{figure}

\begin{figure}[!htbp]
  \centering
  \includegraphics[width=0.4\textwidth]{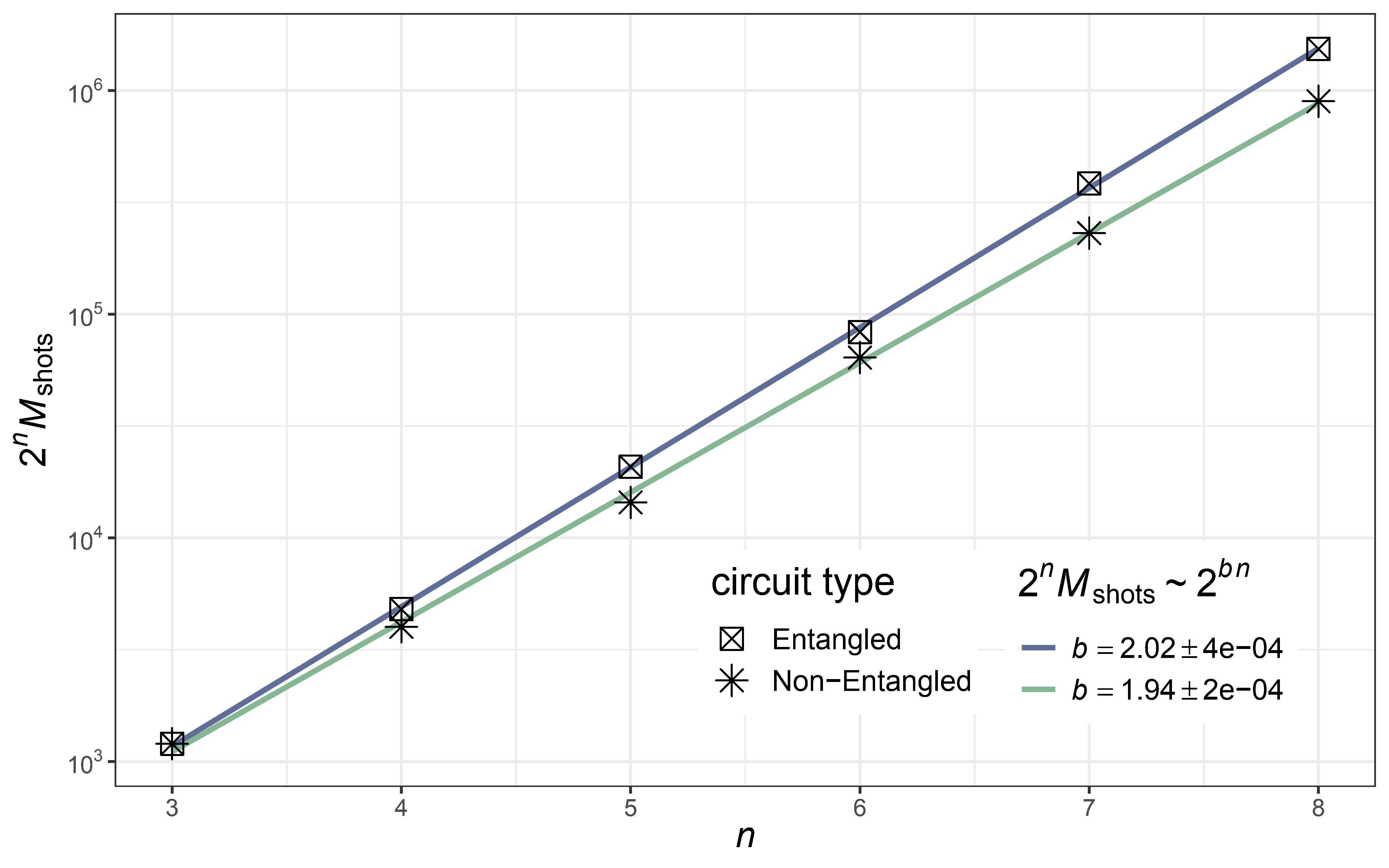}
  \caption{\raggedright
          Scaling of the minimal number of required experimental runs $2^n M_{\rm shots}$ to estimate $\widetilde{F}_{\max}$
          up to a fixed statistical error of $0.05$ as a function of the number of qubits $n$.
          The number of random unitaries is fixed to $N_U=100$.
          The target quantum process is taken to be the $n$-qubit GHZ state preparation circuit for the entangled case 
          and the rotation circuit composed of $n$ single-qubit rotation gates for the non-entangled case.
          The data is obtained via numerical simulation.}
  \label{fig: experiment-2-1}
\end{figure}

In practice, the accuracy of the estimated fidelity is unavoidably subject to statistical error, 
as a result of the finite number of random circuits ($2^n \cdot N_U$) and
the finite number of projective measurements ($M_{\rm shots}$) performed per random circuit.
Therefore, it is experimentally crucial to consider the scaling of the total number of experimental runs, 
which equals to $2^n\cdot N_U\cdot M_{\rm shots}$, constituting the measurement budget, 
in order to effectively suppress the statistical error to a prespecified threshold $\epsilon$
when evaluating the performance of an $n$-qubit quantum process.
In the following, we present numerical simulation to investigate this behavior.

In Figure~\ref{fig: experiment-2}, numerical results for the average statistical error 
as a function of the measurement budget $2^n\cdot N_U \cdot M_{\rm shots}$ are presented  
and the scaling of the measurement budget with respect to the system size $n$ is derived. 
In order to keep consistent with previous experiments, 
we choose the \texttt{H} gate when $n=1$ and the \texttt{CNOT} gate when $n=2$ in the simulation.
Note that in this case the ideal fidelity $F_{\max}=1$ is known.
We repeat our protocol on the ideal simulator $5$ times for each point in the figure 
and record the mean of the statistical errors $|\widetilde{F}_{\max} - 1.0|$. 
We find that the statistical error scales as $|\widetilde{F}_{\max} - 1.0|\sim 1/(2^n N_U M_{\rm shots})$, 
where $\widetilde{F}_{\max}$ is the estimated max process fidelity via simulation.

Now we investigate the scaling of the required number of experimental runs, $2^n M_{\rm shots}$, per unitary to 
estimate the max fidelity $\widetilde{F}_{\max}$ within an average statistical error of $\epsilon = 0.05$ while fixing $N_U$ to $100$.
We employ our protocol to two very different types of quantum processes, with different numbers of qubits $n$: 
(i) a highly entangled quantum process corresponding to an $n$-qubit GHZ state preparation circuit (Entangled) 
and (ii) a completely local quantum process composed of $n$ single-qubit rotation gates (Non-Entangled).
The numerical results are presented in Figure~\ref{fig: experiment-2-1}. 
From the fitted data, we find that $2^nM_{\rm shots}\sim 2^{bn}$, 
where $b = 2.02\pm 4{\rm e-}4$ for the entangled case and $b = 1.94\pm 2{\rm e-}4$ for the non-entangled case.
The analysis shows that our ancilla-free cross-platform protocol requires 
a total number of experimental runs that scales as $2^n N_U M_{\rm shots} \sim 2^{b n}$ with $b \approx 2$. 
This scaling, despite exponential, is significantly less 
than full quantum process tomography (QPT), which has an exponent $b \geq 4$~\cite{mohseni2008quantum}.

\section{Conclusions}\label{sec:conclusions}

We have proposed an ancilla-free cross-platform protocol that enables the performance comparison of arbitrary quantum processes,
using only single-qubit unitaries and classical communication.
This protocol is thus suitable for comparing quantum processes that are independently manufactured over different times and locations, 
built by different teams using different technologies.
We have experimentally demonstrated the cross-platform protocol on six remote quantum computers fabricated by IBM and Baidu,
and monitored the stable functioning of Baidu's "Qianshi" quantum computer over one week.
The experimental results reveal that our protocol accurately compares 
the performance of different quantum computers with significantly fewer measurements than quantum process tomography.  
Additionally, we have shown that our protocol is applicable to quantum process tomography. 

However, some problems must be further explored to make the cross-platform protocols more practical.
Firstly, the sample complexity of these protocols lacks theoretical guarantees, 
thereby necessitating the empirical selection of experimental parameters. 
To address this challenge, it may be possible to adapt techniques from~\cite{Anshu2022}.
Secondly, it is vital to make the protocols robust against state preparation and measurement errors.
One possible solution is to apply quantum error mitigation methods~\cite{temme2017error,kandala2019error,wang2021mitigating,tang2022detecting,cai2022quantum} 
to alleviate quantum errors and increase the estimation accuracy.
We suggest that ideas and insights from randomized benchmarking~\cite{PRXQuantum.3.020357} and 
quantum gateset tomography~\cite{Nielsen2021gatesettomography} might be helpful
for designing error robust cross-platform protocols.

\section*{Acknowledgements}

Part of this work was done when C. Z. was a research intern at Baidu Research.
We acknowledge the use of IBM Quantum services for this work. 
The views expressed are those of the authors, and do not reflect the official policy or position of 
IBM or the IBM Quantum team.
This work was partially supported by the National Science Foundation of China (Nos. 61871111 and 61960206005). 


%

\begin{appendices}
\onecolumngrid

\section{Proof of the protocol}\label{appx: proof of cpe for process}

In this section, we prove the correctness of the ancilla-free cross-platform comparison protocol introduced in the main text (MT). 

For two $n$-qubit quantum processes $\cE_1$ and $\cE_2$ performed on platforms $\cS_1$ and $\cS_2$, 
let's start with the ancilla-assisted idea for estimating $\tr[\eta_i\eta_j]$. 
On each platform, we prepare a $2n$-qubit maximally entangled state with an $n$-qubit auxiliary system 
and apply the corresponding quantum process $\cE_i$ to the target quantum system. 
This enables us to construct the Choi state $\eta_i$ of the quantum process $\cE_i$. 
Using Eq.~\eqref{eq: cpe for state} in MT, we can estimate the overlap between the Choi states $\eta_i$ and $\eta_j$, where $i,j=1,2$, via 
\begin{align}
  \tr[\eta_i\eta_j] = 4^n\sum_{\bm{s}, \bm{s}', \bm{k}, \bm{k}'\in\{0,1\}^{n}} 
        (-2)^{-\cD[\bm{s}, \bm{s}'] -\cD[\bm{k}, \bm{k}']} \overline{{\rm Pr}_{U}^{(i)}[\bm{s},\bm{k}]{\rm Pr}_{U}^{(j)}[\bm{s}',\bm{k}']},
        \label{eq:ancilla-assisted}
\end{align}
where $\bm{s}$ and $\bm{k}$ are bitstrings representing the outcomes of the auxiliary system and the original system, respectively.
$\overline{\cdots}$ denotes ensemble average over $U$, where $U$ is a tensor product of $2n$ local random unitaries 
sampled from a single-qubit set $\cX_2$ satisfying unitary $2$-design. 

The key trick for removing the ancilla qubits from~\eqref{eq:ancilla-assisted} is that we can generate the probability 
distribution ${\rm Pr}_{U}^{(i)}[\bm{s},\bm{k}]$ through a prepare-and-measure approach, 
without requiring the use of any ancilla qubits.
Since $U$ is a tensor product of $2n$ local random unitaries, we can decompose it as 
$U=U_1\otimes U_2$, where $U_1$ and $U_2$ are the unitaries operating on the auxiliary and original systems, respectively.
Note that both $U_1$ and $U_2$ are tensor products of $n$ local random unitaries. 
Based on this decomposition, we can rewrite the probability 
distribution ${\rm Pr}_{U}^{(i)}[\bm{s},\bm{k}]$ in Eq.~\eqref{eq:ancilla-assisted} as
\begin{align}
      {\rm Pr}_{U}^{(i)}[\bm{s},\bm{k}]
&=   \tr\left[(\ketbra{\bm{s}\bm{k}}{\bm{s}\bm{k}})\left(U \eta_i U^\dagger\right)\right] \\
&=   \tr\left\{(\ketbra{\bm{s}\bm{k}}{\bm{s}\bm{k}})\left[(U_1\otimes U_2)(\cI\ox\cE_i)(\ketbra{\psi_+}{\psi_+}) (U_1\otimes U_2)^\dagger\right]\right\} \\
&=   \tr\left\{\left[(U_1^\dagger\ketbra{\bm{s}}{\bm{s}}U_1)\ox\left( 
        (\ketbra{\bm{k}}{\bm{k}})(\cU_2\circ\cE_i)\right)\right](\ketbra{\psi_+}{\psi_+})\right\} \\
&=   \tr\left\{(U_1^\dagger\ketbra{\bm{s}}{\bm{s}}U_1)\ox
        \ketbra{\bm{k}}{\bm{k}}(\cU_2\circ\cE_i)(U_1^T\ketbra{\bm{s}}{\bm{s}}U_1^*)\right\}\label{eq:tmp3} \\
&=   \tr\left[\ketbra{\bm{k}}{\bm{k}}(\cU_2\circ\cE_i)(U_1^T\ketbra{\bm{s}}{\bm{s}}U_1^*)\right], \label{eq: formula of protocol}
\end{align}
where $\cU_i(\cdot) = U_i (\cdot) U_i^\dagger$ and Eq.~\eqref{eq:tmp3} follows from the identity $(\ketbra{u}{u}\ox\1)\ket{\psi_+} = \ket{u}\ox\ket{u^*}$. 
Eq.~\eqref{eq: formula of protocol} leads to an ancilla-free approach to generate the probability 
distribution ${\rm Pr}_{U}^{(i)}[\bm{s},\bm{k}]$ as follows:
We prepare the input state $U_1^T\ket{\bm{s}}$, execute the target quantum process $\cE_i$, 
followed by the unitary $U_2$, and measure the resulting state in the computational basis.
The remaining question is what probability distribution should be used to sample and prepare the computational basis input states $\ket{\bm{s}}$.
Since the reduced state of a maximally entangled state is the maximally mixed state, 
we conclude that the measurement outcomes of the auxiliary system satisfy the uniform distribution for arbitrary randomized measurements, i.e.,
\begin{align}
    \Pr\left\{S = \bm{s} | U_1\right\} = \frac{1}{2^n}.
\end{align}
That is, the computational basis input states $\ket{\bm{s}}$ should be sampled and prepared with respect to the uniform distribution.
This accomplishes the final piece of the ancilla-free cross-platform estimation protocol. 

\section{Randomized quantum process tomography}\label{appx: proof of randomized qpt}

In this section, we prove the correctness of randomized quantum process tomography in the MT.

In \cite{elben2019statistical}, it was shown that the full information of an $n$-qubit unknown quantum state $\rho$ can be 
reconstructed using the experimental data $\{\Pr_U^{(1)}, \Pr_U^{(2)}\}_U$ collected in Section~\ref{sec: state fidelity} via
\begin{align}
    \rho_i = 2^n \sum_{\bm{s}, \bm{s}'\in\{0,1\}^n}(-2)^{-\cD[\bm{s}, \bm{s}']} \overline{{\rm Pr}_{U}^{(i)}[\bm{s}]U^\dagger\ketbra{\bm{s}'}{\bm{s}'}U}, 
    \label{eq:appx:randomized qst}
\end{align}
where $i=1,2$ and $\overline{\cdots}$ denotes ensemble averaging over $U$, 
which is a tensor product of $n$ local random unitaries sampled from $\cX_2$. 

Similarly, we can adopt the experimental data collected by cross-platform process comparison protocols 
to reconstruct the full information of an unknown quantum process. 
Let's begin with the ancilla-assisted protocol for reconstructing an $n$-qubit quantum process $\cE_i$, $i=1,2$.
We prepare a $2n$-qubit maximally entangled state with an $n$-qubit auxiliary system. 
Then the randomized measurement protocol outlined in~\eqref{eq:appx:randomized qst}
provides a simple way to reconstruct the Choi state of the quantum process $\cE$ via
\begin{align}
\eta_{\cE_i} = 4^n \sum_{\bm{s}, \bm{s}', \bm{k}, \bm{k}'\in\{0,1\}^n}
        (-2)^{-\cD[\bm{s}, \bm{s}']-\cD[\bm{k}, \bm{k}']} \overline{{\rm Pr}_{U}^{(i)}[\bm{s}, \bm{k}]U^\dagger
        \ketbra{\bm{s}'\bm{k}'}{\bm{s}'\bm{k}'}U}, \label{eq: formula of rm qpt}
\end{align}
using the experimental data ${\rm Pr}_{U}^{(i)}[\bm{s}, \bm{k}]$ collected in Section~\ref{sec:ancilla-free}.
Note that here $\overline{\cdots}$ denotes ensemble averaging over $U$, which is a tensor product of $2n$ local random unitaries sampled from $\cX_2$. 

As proved in Appendix~\ref{appx: proof of cpe for process}, 
the probability distribution ${\rm Pr}_{U}^{(i)}[\bm{s},\bm{k}]$ can be generated using a prepare-and-measure 
approach that does not require the use of ancilla qubits. 
Specifically, we decompose each $U$ as $U=U_1\otimes U_2$, 
where $U_1^T$ and $U_2$ are employed for state preparation and randomized measurement, respectively.
The ancilla-free approach reconstructs the Choi state of the quantum process $\cE$ 
with the same formula as Eq.~\eqref{eq: formula of rm qpt}.
The advantage is that it facilitates the reconstruction in a more practical manner.

\end{appendices}

\end{document}